\documentclass[%
 reprint,
superscriptaddress,
 amsmath,amssymb,
]{revtex4-2}

\usepackage{natbib}

\makeatletter
\def\NAT@def@citea{\def\@citea{\NAT@separator}}
\makeatother

\usepackage{graphicx}
\usepackage[colorlinks,allcolors=blue]{hyperref}
\usepackage{dcolumn}
\usepackage{setspace}

\begin{document}

\title{Programmable heralded linear optical generation of two-qubit  states}

\author{Suren A.\,Fldzhyan}
\affiliation{%
 Quantum Technology Centre and Faculty of Physics, M.V. Lomonosov Moscow State University, 1 Leninskie Gory Street, Moscow 119991, Russian Federation
}%
\affiliation{%
 Russian Quantum Center, Bolshoy bul'var 30 building 1, Moscow 121205, Russian Federation
}%
\author{Mikhail Yu.\,Saygin}%
 \email{saygin@physics.msu.ru}
\affiliation{%
 Quantum Technology Centre and Faculty of Physics, M.V. Lomonosov Moscow State University, 1 Leninskie Gory Street, Moscow 119991, Russian Federation
}%
\affiliation{Laboratory of Quantum Engineering of Light, South Ural State University (SUSU), Russia, Chelyabinsk, 454080, Prospekt Lenina 76}%
\author{Sergei P.\,Kulik}
\affiliation{%
 Quantum Technology Centre and Faculty of Physics, M.V. Lomonosov Moscow State University, 1 Leninskie Gory Street, Moscow 119991, Russian Federation
}%
\affiliation{Laboratory of Quantum Engineering of Light, South Ural State University (SUSU), Russia, Chelyabinsk, 454080, Prospekt Lenina 76}%


\begin{abstract}

We investigated the heralded generation of two-qubit dual-rail-encoded states  by programmable linear optics. Two types of scheme generating the states from four single photons, which is the minimum possible to accomplish the task, were considered. The schemes have different detection patterns heralding successful generation events; namely,  one-mode heralding, in which the two auxiliary photons are detected in one mode, and two-mode heralding, in which single photons are detected in each of the two modes simultaneously. We show that the dependencies of the schemes' success probabilities on the target state's degree of entanglement are essentially different. In particular, one-mode heralding yields greater efficiency for highly entangled states if the programmable interferometers can explore the full space of the unitary transfer matrices. The reverse is the case for weakly entangled states where two-mode heralding is better. We found a minimal decomposition of the scheme with two-mode heralding that is programmed by one variable phase shift. We infer that the linear optical schemes designed specifically for the generation of two-qubit states are more efficient than schemes implementing known linear optical gate--based circuits. Our results reveal a substantial reduction of physical resources needed to generate some types of nonmaximally entangled multiqubit states used in quantum computing.

\end{abstract}

\maketitle

\section{Introduction}

Photonics plays an indispensable role in quantum technologies, because it offers convenient approaches to  processing and transmitting quantum information~\cite{QuantumPhotonicsReview,PhotonicQIPreview}.  In particular, optical photons are ideal carriers that enable the transmission of quantum information between distant parties  through established fiber lines ~\cite{EntanglementQKD,EntangledNetwork}. The qubits encoded into flying photons and manipulated by conventional photonic devices can also be used in realizations of quantum computers, which is currently a strong contender platform in quantum computing ~\cite{Carolan,FBcomputingPsi,XanaduChip}. 

Linear optics can generate entangled states out of separable single photons and perform some transformations on the states~\cite{aaronson2010,BosonSampling20photons}. However, specific states that encode logical qubits, which are usually required by quantum algorithms, are challenging to obtain by linear optics. Most often, a photonic qubit is encoded by a single photon being in either of two modes, as both the loss and multi photon contamination can be easily detected  by measuring the total photon number in the two modes. These so-called dual-rail-encoded states are considered throughout this paper.  

Some limitations of linear optics are manifested in probabilistic generation of the states. Moreover, linear optical generation usually requires extra photons in addition to the ones encoding the qubits themselves~\cite{ResourceCost,stanisic,fldzhyan2021compact,Carolan,LaingGHZ}. Typically, the  success probability of a linear optical scheme rapidly diminishes with increasing dimension of the entangled qubit states~\cite{gubarev2020,fldzhyan2021compact,gubarev2020,Blasiak2022}. Fortunately, there is a quantum computing model using multiqubit maximally entangled cluster states, constructed from many identical small-scale entangled states, such as two-qubit Bell states and three-qubit Greenberger-Horne-Zeilinger (GHZ) states, that serve as a universal computational resource~\cite{MBQC,MBVQE,OptimisticRudolph}. With cluster states at hand, a specific quantum algorithm is then implemented by the performing of single-qubit measurements and single-qubit operations that are deterministic in linear optics. Recently, a more-advanced model of quantum computation, called ``fusion-based quantum computing," was proposed, in which computation is done by entangling measurements of small-scale entangled states without the generation of a large universal cluster state before implementation of the actual algorithm as in measurement-based quantum computing~\cite{FBcomputingPsi}. In both models, the quantum computation is fueled by the sources of the small-scale entangled states, so achieving high generation efficiency of such states is important for photonic quantum computers.

Although the generation of few-qubit entangled states is still nondeterministic, it can now be overcome by multiplexing multiple schemes and using them in parallel or by consecutive iterations of a single scheme, enabling near-deterministic generation. To gain an advantage from multiplexing, one requires the heralding of successfully generated states without destroying them by a measurement. At the same time, one needs fast and low-loss feed-forward operations for active switching of states produced by a random scheme in order to send it to a predefined set of modes~\cite{ResourceCost,PSIentangledstates}. As a result, large-scale schemes are greedy with regard to the volume and the quality of physical resources. It becomes very challenging to devise a multiplexing circuit that meets such stringent requirements. Crucially, the scale and the complexity of passive and active parts of the circuits are dependent on the success probability with which the schemes operate, providing motivation for the development of more-efficient linear optical approaches to the generation of few-qubit entangled qubit states.

Recently, linear optical generation of maximally entangled dual-rail Bell states and dual-rail three-qubit GHZ states was thoroughly investigated~\cite{stanisic,gubarev2020,Gubarev2021,fldzhyan2021compact,PSIentangledstates}. In particular, it has been shown that at least four photons and six photons, respectively, are required to generate such states. It has been shown that the generation efficiency of Bell states by four-photon schemes depends on the measurement pattern used to herald successful generation events~\cite{fldzhyan2021compact}. 

Nonmaximally entangled states can also serve as a resource for quantum computation. In particular, weighted-graph states, which can be considered as a generalization of the traditional maximally entangled cluster states, have potential uses in measurement-based quantum computation and self-correcting quantum memories~\cite{XP_stabilizers}. Therefore, it is of applied interest to investigate methods for creating such states, from the problem of linear optical generation of small-scale nonmaximally entangled states to their fusion into large multiqubit states. This work aims to investigate the former problem, while the latter will be considered in other work~\cite{our_future_work}. Namely, we study the efficiency of linear optics to produce generic two-qubit dual-rail-encoded states with an arbitrary degree of entanglement. Although these states are the simplest possible, they can potentially fuel the generation of multiqubit-resource states, similarly to the cluster-based or fusion-based quantum computing. 

This paper is arranged as follows. In Sec.~\ref{sec:ParametrizedStates} we introduce two programmable optical schemes for the generation of the arbitrary two-qubit states studied. In Sec.~\ref{sec:GateBasedGeneration} the generation of the two-qubit states by programmable gate--based quantum circuits, including their known linear optical implementations, is described. In Sec.~\ref{sec:OptimizationMethod} we introduce the methods that we used to study the efficiency the schemes can offer  and to design optical schemes. The results obtained are presented in Sec.~\ref{sec:results}. We summarize and discuss the results in Sec.~\ref{sec:discussion}.

\section{Two-qubit entangled states}\label{sec:ParametrizedStates}

We are interested in linear optical generation of two-qubit states in the dual-rail encoding. In this encoding a qubit state is a superposition of a single photon over two modes. Namely, two  basis states of a logical qubit, $|0\rangle_L$ and $|1\rangle_L$, are encoded into the states $|0\rangle_m|1\rangle_n$ and $|1\rangle_m|0\rangle_n$ carried by modes with indices $m$ and $n$, where $|n\rangle_j$ is the $n$-photon Fock state of mode $j$. In this encoding any single-qubit operation can be implemented deterministically and with high accuracy by linear optical elements. However, this is not the case for two-qubit operations generating entangled states, which are generally nondeterministic and require extra  photons. An arbitrary two-qubit state $|\Psi_{2q}\rangle=c_{00}|0\rangle_{L1}|0\rangle_{L2}+c_{11}|1\rangle_{L1}|1\rangle_{L2}+c_{01}|0\rangle_{L1}|1\rangle_{L2}+c_{10}|1\rangle_{L1}|0\rangle_{L2}$ can be reduced to the canonical form $|\Psi_{2q}'\rangle=a|0\rangle_{L1}|0\rangle_{L2}+b|1\rangle_{L1}|1\rangle_{L2}$ by suitable single-qubit operations. With this in mind, we consider the linear optical generation of the following dual-rail-encoded states (for brevity, mode designations are omitted):
    \begin{equation}\label{eqn:parametrized_state}
        |\Phi(\alpha)\rangle=\cos\alpha|1010\rangle+\sin\alpha|0101\rangle,
    \end{equation}
where  $a=\cos\alpha$ and $b=\sin\alpha$ are parametrized by angle $\alpha$.  One can quantify the entanglement degree in any of the convenient measures, for example, the von Neumann entropy $S(\hat{\rho}_j)=-\text{Tr}(\hat{\rho}_j\log_2\hat{\rho}_j)=-(a^2\log_2a^2+b^2\log_2b^2)=-\left(\cos^2\alpha\log_2(\cos^2\alpha)+\sin^2\alpha\log_2(\sin^2\alpha)\right)$, where the reduced density 
$\hat{\rho}_j=\text{Tr}_{2-j}(|\Phi(\alpha)\rangle\langle\Phi(\alpha)|)$ ($j=1,2$), with $\text{Tr}_i(\cdot)$ being the trace with respect to the subsystem $i$. However, in the following we use the angle parameter $\alpha$ to quantify the entanglement degree. It takes values in the range from $0$ to $\pi/4$, where the bounds correspond to separable states and the maximally entangled Bell state, respectively.

In general, unitary evolution of single-photon states through interferometers provides only a limited class of entangled states. Therefore, one cannot generate an arbitrary entangled state by static linear optics. To broaden the class of accessible states, one can prepare entangled states carried by a larger number of modes and photons than those of the target qubit states. These states are outside the Hilbert space corresponding to desired logical qubit encoding. Then, some modes of the states are measured nondeterministically, collapsing the unmeasured modes into the target states. In particular, this approach has been applied for the generation of two-qubit dual-rail-encoded Bell states~\cite{fldzhyan2021compact,gubarev2020,stanisic} and the three-qubit GHZ state~\cite{gubarev2020,stanisic}. It has been shown that at least four single photons are required to generate Bell states by linear optics---two of which are to be measured to herald successful generation~\cite{fldzhyan2021compact,gubarev2020,stanisic}. In this work, we consider a more-general task of nondeterministic generation of states \eqref{eqn:parametrized_state} using lossless multiport interferometers and four indistinguishable photons.

    \begin{figure}[htp]
        \centering
        \includegraphics[width=0.45\textwidth]{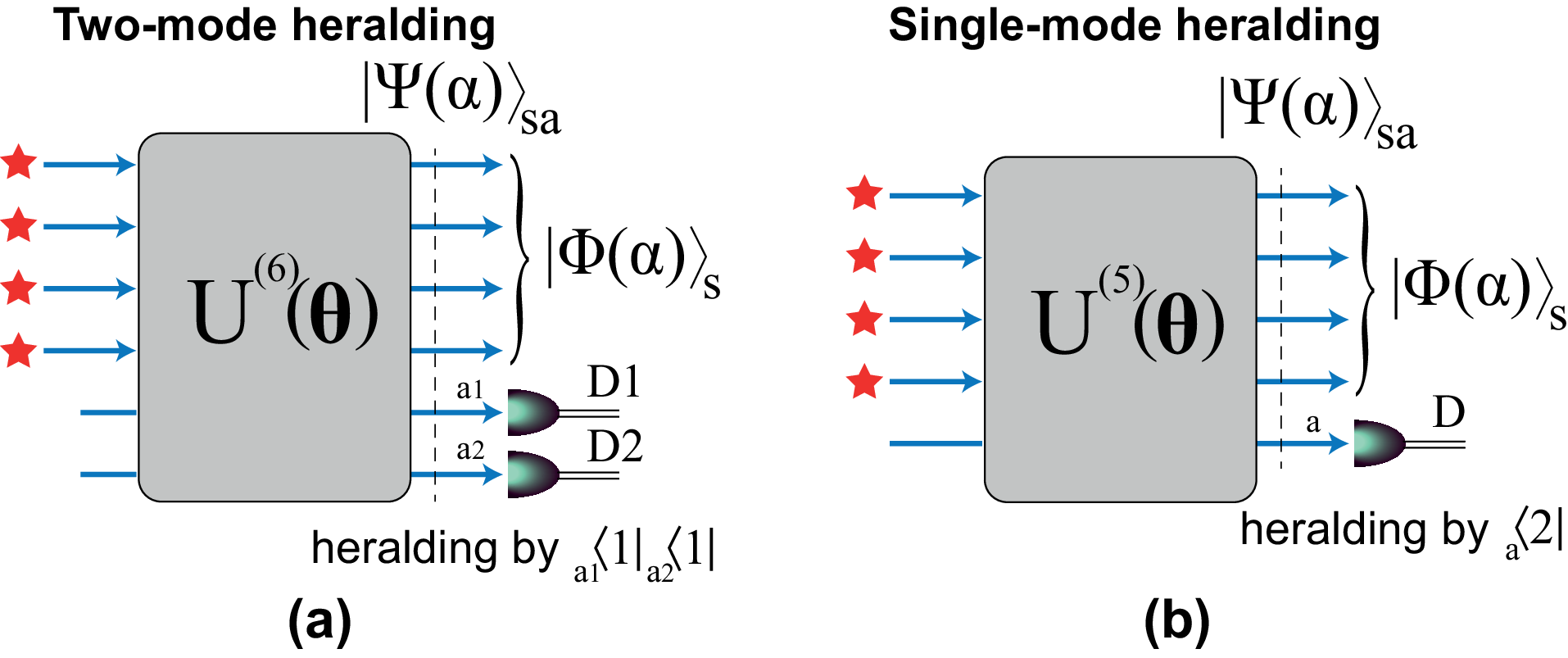}
        \caption{Two types of linear optical scheme considered for the generation of two-qubit states \eqref{eqn:parametrized_state}. a) Scheme with two-mode heralding:  detectors D1 and D2 should measure single photons simultaneously in both auxiliary modes a1 and a2 for heralding. b) Schemes with single-mode heralding: photon detector D should measure two photons in auxiliary mode a for heralding. A specific generated state $|\Phi(\alpha)\rangle$ is defined by the reprogramming of transfer matrices $U^{(6)}$ and $U^{(5)}$ of the multiport interferometers through proper setting of the parameters $\boldsymbol{\theta}$. }
        \label{fig:fig_1}
    \end{figure}

Even though the heralding efficiency is typically much less than unity, one knows exactly when successful events occur, as they are flagged by a pattern of photons that are detected in auxiliary modes. This property makes the heralded nondeterministic schemes practical, in contrast to postselection schemes, where the success is verified only by measuring the generated state~\cite{Kieling_2010,CNOT_coincidence_basis}. Assuming that success is heralded by the detection of a pattern of photons $\mathbf{d}$ in the auxiliary modes, the output quantum state prepared by the interferometers should take the following form:
    \begin{equation}\label{eqn:optimal_state_form}
        |\Psi(\alpha)\rangle_{sa}=\sqrt{p(\alpha)}|\Phi(\alpha)\rangle_s|\mathbf{d}\rangle_a+\sqrt{1-p(\alpha)}|R(\alpha)\rangle_{sa},
    \end{equation}
where $p(\alpha)$ is the success probability of preparing the target state $|\Phi(\alpha)\rangle$, which we expect to depend on the state parameter $\alpha$, and $|R(\alpha)\rangle_{sa}$ is the residual component that accompanies the linear optical generation. In Eq.~\eqref{eqn:optimal_state_form} we mark the output modes that carry the target states and the auxiliary (heralded) modes by indices ``s" and ``a," respectively. Obviously, the detection pattern $\mathbf{d}$ should obey $_a\langle\mathbf{d}|R\rangle_{sa}=0$ to project the large state $|\Psi(\alpha)\rangle_{sa}$ unambiguously onto the smaller target state $|\Phi(\alpha)\rangle$, i.e.  $_a\langle\mathbf{d}|\Psi(\alpha)\rangle_{sa}\propto|\Phi(\alpha)\rangle_s$, the residual component.

We study two types of compact linear optical scheme, as shown in Fig.~\ref{fig:fig_1}. The schemes use four indistinguishable photons and differ by the measurement pattern used to herald successful generation events: (1) the scheme shown in Fig.~\ref{fig:fig_1}(a) uses a six-by-six multiport interferometer and two-mode heralding measurement, in which two photodetectors, D1 and D2, measure auxiliary modes labeled as ``a1" and ``a2"; (2) the scheme shown in Fig.~\ref{fig:fig_1}(b) uses a smaller (five-by-five) multiport interferometer and it relies on single-mode heralding and needs only one photodetector, D. The states produced by these two schemes are heralded by the measured patterns $|\mathbf{d}\rangle_a=|1\rangle_{a1}|1\rangle_{a2}$ and $|\mathbf{d}\rangle_a=|2\rangle_{a}$, respectively. We denote these schemes as the ``two-mode-heralded sceme" and the ``single-mode-heralded scheme," respectively.

\section{Gate-based linear optical circuits}\label{sec:GateBasedGeneration}

\begin{figure}[htp]
        \centering
        \includegraphics[width=0.4\textwidth]{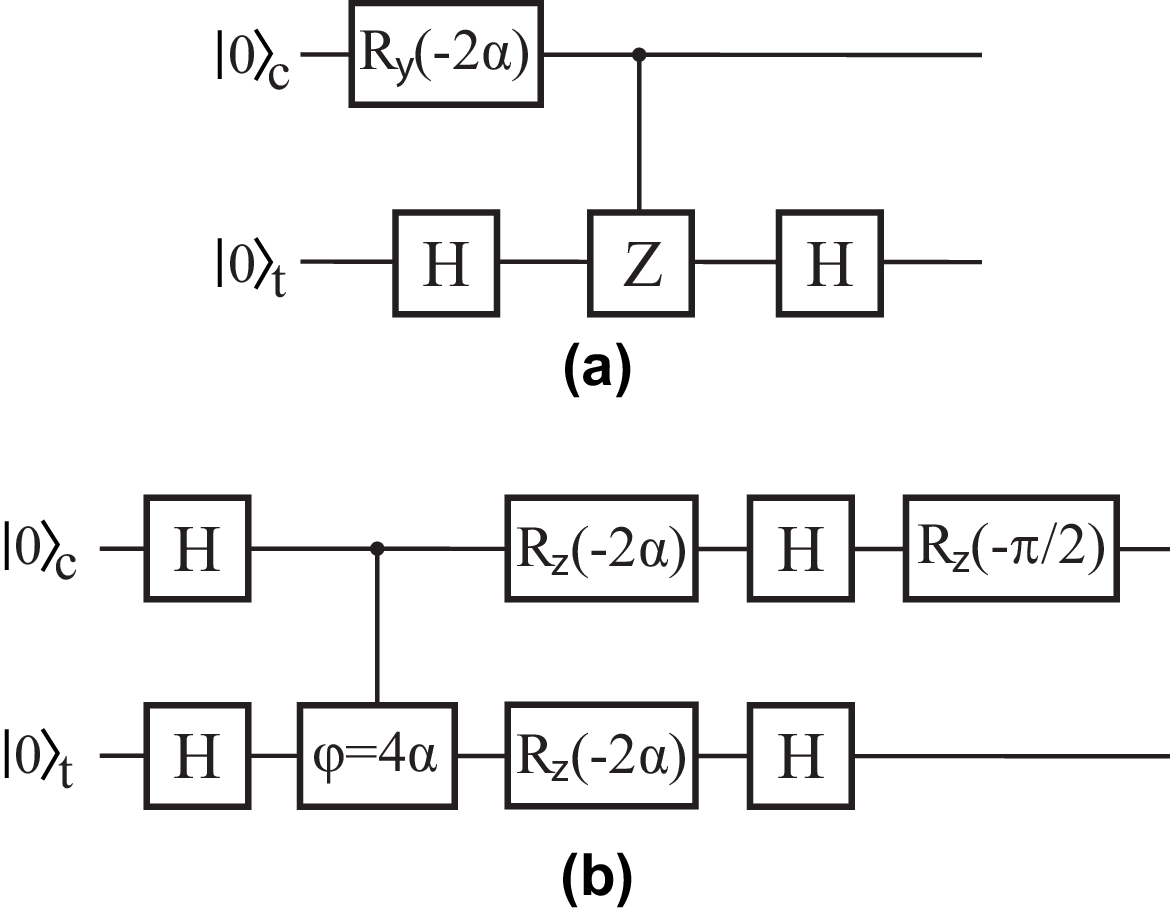}
        \caption{Programmable quantum circuits for the generation of entangled two-qubit states $|\Phi(\alpha)\rangle$. (a) A circuit using a two-qubit {\scriptsize CZ} gate and variable single-qubit rotation $\hat{R}_y(-2\alpha)$. (b) A circuit using a two-qubit {\scriptsize CPHASE}($\varphi$) gate with variable parameter $\varphi$ that controls the degree of entanglement given by $\varphi=4\alpha$. The single-qubit gates applied after the two-qubit gates are needed solely to reducing the state to its canonical form \eqref{eqn:parametrized_state}. Other gates in the circuits are a Hadamard gate, $\hat{H}$, and rotation around the $z$ axes on the Bloch sphere, $\hat{R}_z(\varphi)=\exp\left(-i\varphi/2\hat{\sigma}_z\right)$, where $\hat{\sigma}_z$ is the Pauli operator.}
        \label{fig:fig_2}
    \end{figure}

To compare the effectiveness of the two schemes shown in Fig.~\ref{fig:fig_1}, we first consider previously known approaches to accomplish the task by linear optics. To the best of our knowledge there are only two such programmable quantum circuits that can generate the arbitrary two-qubit state shown in Fig.~\ref{fig:fig_2}. A specific state $|\Phi(\alpha)\rangle$ is produced by the setting of proper values of the gate's variable parameters. The basic difference between these circuits is in the type of entangling gate used. Namely, the circuit shown in Fig.~\ref{fig:fig_2}(a) produces entanglement between the qubits by the static controlled $Z$ ({\scriptsize CZ}) gate while a variable single-qubit gate performs rotation around the $y$ axis of the Bloch sphere,
    \begin{equation}\label{eqn:Ry}
        \hat{R}_y(\varphi)=\exp\left(-i\frac{\varphi}{2}\hat{\sigma}_y\right),
    \end{equation}
which is applied to the control qubit before the {\scriptsize CZ} gate. $\varphi$ is a variable parameter and $\hat{\sigma}_y$ is the Pauli operator. $\varphi$ defines the degree of entanglement for the generated state: to generate state $|\Phi(\alpha)\rangle$, one sets $\varphi=-2\alpha$. Notice that the entangling gate in this circuit is static, i.e. it performs transformations irrespective of the target state.  

Fig.~\ref{fig:fig_2}(b) shows another approach to two-qubit-state generation, in which the variable controlled-phase {\scriptsize CPHASE}-gate is used to entangle the qubits. Here, the entanglement degree is controlled directly by the two-qubit gate rather than by the variable single-qubit rotation as in the circuit in Fig.~\ref{fig:fig_2}a. This gate performs the following transformation of logic qubits:
    \begin{equation}\label{eqn:cphase_matrix}
        \text{\scriptsize CPHASE}(\varphi)=\text{diag}(1,1,1,e^{i\varphi}),
    \end{equation}
where $\varphi$ is a variable parameter taking values in the range from $0$ to $\pi$, corresponding to the identity (nonentangling) and the {\scriptsize CZ} operation, respectively. To generate $|\Phi(\alpha)\rangle$, one sets the gate parameter $\varphi$ equal to $4\alpha$. 

Ssingle-qubit operations are deterministic in the dual-rail encoding and, therefore, in linear optical implementations the success probability does not change with $\alpha$ in the circuit shown in Fig.~\ref{fig:fig_2}(a). However, as will be clear from the examples that follow, the generation-success probability of the {\scriptsize CPHASE} gate--based circuit shown in Fig.~\ref{fig:fig_2}(b) depends on $\alpha$. 

Let us recall some previous results on linear optical {\scriptsize CZ} and programmable {\scriptsize CPHASE} gates. There is a heralded linear optical controlles {\scriptsize NOT} ({\scriptsize CNOT}) gate requiring four photons, two of which are auxiliary and are used in heralding measurement~\cite{CNOT_heralded}. The {\scriptsize CZ} gate can be transformed into the {\scriptsize CNOT} gate by single-qubit operations.  The success probability of this gate is $1/16$. A higher success probability of $1/9$ is achieved by a circuit that uses a post-selected {\scriptsize CNOT} gate~\cite{CNOT_coincidence_basis}, however, one cannot check successful events without measuring the photons carrying the target state, rendering the gate impractical. 


The heralded {\scriptsize CPHASE} gate can be constructed with two single-mode nonlinear phase shift gates, transforming the states as $c_0|0\rangle+c_1|1\rangle+c_2|2\rangle\rightarrow{}c_0|0\rangle+c_1|1\rangle+c_2e^{i\varphi}|2\rangle$  \cite{Scheel2003}. However, the success probability of linear optical implementations of the gate does not exceed $1/16$ at $\varphi\neq0$\cite{Sparrow2018}, which is lower than for the gate--based circuits used in the analysis below.

\begin{figure}[htp]
        \centering
        \includegraphics[width=0.45\textwidth]{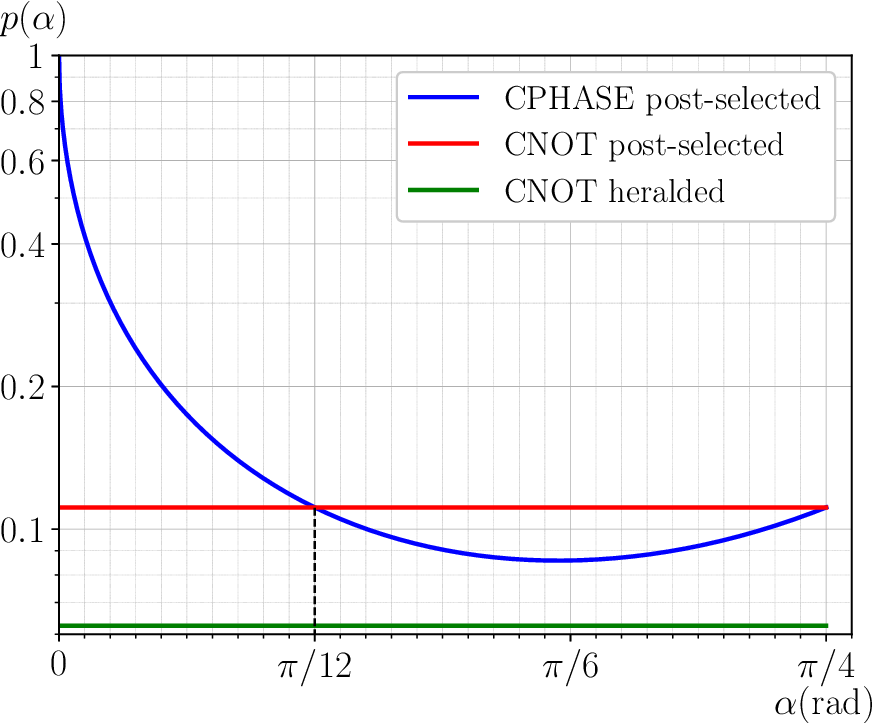}
        \caption{Success probability of the linear optical implementations of the {\scriptsize CNOT}/{\scriptsize CZ} gate~\cite{CNOT_coincidence_basis} and the variable {\scriptsize CPHASE}($\varphi$) gate~\cite{Kieling_2010} as a function of $\alpha=\varphi/4$ that specify the generated state. The post-selected {\scriptsize CNOT} gate has a success probability of $1/9$; the heralded {\scriptsize CNOT} gate has a success probability of $1/16$; the {\scriptsize CPHASE} gate efficiency curve is given by Eq.~\eqref{eqn:cphase_dependence}.}
        \label{fig:fig_3}
    \end{figure}

The most-efficient linear optical implementation of the {\scriptsize CPHASE} gate known today operates with postselection~\cite{Kieling_2010}. It requires no auxiliary photons (only two photons encoding the transformed state enter the circuit). The probability of the post-selected {\scriptsize CPHASE} operation depends on the gate parameter $\varphi$ and is given by~\cite{Kieling_2010}
    \begin{equation}\label{eqn:cphase_dependence}
        p(\varphi)=\left(1+2\sin\frac{\varphi}{2}+2^{3/2}\sin\frac{\pi-\varphi}{4}\sqrt{\sin\frac{\varphi}{2}}\right)^{-2},
    \end{equation}
which is plotted in Fig.~\ref{fig:fig_3} as a function of generated-state parameter $\alpha=\varphi/4$. As can be seen from the Fig.~\ref{fig:fig_3}, the gate operates most effectively at small values of $\alpha$, becoming deterministic at $\alpha\rightarrow{}0$. Notice that at $\alpha\in[\pi/12,\pi/4)$ the efficiency of the {\scriptsize CPHASE} gate--based scheme is lower than of the static {\scriptsize CZ} gate--based one. Therefore, if one takes efficiency into account, the choice between the two linear optical circuits depends on the specific target state: if $\alpha\in[0,\pi/12]$, the circuit based on the {\scriptsize CPHASE} gate performs better, whereas if $\alpha\in[\pi/12,\pi/4)$, the {\scriptsize CZ} gate--based circuit is better. In what follows we will compare our results with results corresponding to the most effective post-selected implementation of the {\scriptsize CZ} and {\scriptsize CPHASE} gates, as well as to the less-effective heralded {\scriptsize CZ} gate. We understand, however, that practical usage of the post-selected circuits is dubious.


\section{Design of programmable linear optical circuits}\label{sec:OptimizationMethod}

For practical systems the transfer matrices of the multiport interferometers should be reprogrammed each time a new target state is required. We study different designs of programmable multiport interferometers to achieve as high a success probability $p(\alpha)$ as possible,  while keeping the photonic circuits as compact as possible. For this, we investigate two possible interferometer designs that differ by the number of variable phase shifts and, hence, by the complexity of an experimental realization. Namely, we investigate universal and restricted programmable interferometers.

\subsection{Universal linear optics}

First, we consider the case of universal interferometers, i.e., ones that are capable of implementing arbitrary transfer matrices  ($U^{(6)}$ and  $U^{(5)}$ in Fig.~\ref{fig:fig_1}). Therefore, universal interferometers allow a full exploration of the capabilities of linear optics to generate entangled qubit states. 

These interferometers usually come in the form of two-mode beam-splitter meshes having variable splitting ratios and phase shifts.  In particular, we use the rectangular design proposed in Ref.~\cite{Clements2016}. According to this design, a universal decomposition represents an $N$-mode interferometer as $N$ layers consisting of beam-splitters with variable transmissivities and variable phase shifts. Each beam-splitter acts locally on two neighboring modes by a transfer matrix (see Appendix \ref{appendix:parametrization}):
    \begin{equation}\label{eqn:BSmatrix}
       T_j(\theta',\theta'')=\left(
        \begin{array}{cc}
             e^{i\theta''}\sin\theta' & \cos\theta' \\
             e^{i\theta''}\cos\theta' & -\sin\theta'
        \end{array}\right),
    \end{equation}
where $\theta'$ specifies the transmissivity of the beam-splitter $\tau=\cos^2\theta'$ ($0\le\theta'\le\pi/2$), and $\theta''$ is the phase difference between the input modes of the beam-splitter ($-\pi\le\theta''<\pi$). The resultant multimode transfer matrix of the universal interferometer is the product of $N(N-1)/2$ matrices: $U=D\times T_Q^{(n_Q,m_Q)}\times\ldots\times T_1^{(n_1,m_1)}$, with $D$ being a diagonal matrix of independent $N-1$ phase shifts $\theta'''_j$ and $T^{(m_q,n_q)}_q$ being the beam-splitter matrix acting on modes $m_q$ and $n_q$. 

To analyze the capabilities of the universal schemes in the generation of two-qubit states \eqref{eqn:parametrized_state}, we apply the limited-memory Broyden-Fletcher-Goldfarb-Shanno optimization algorithm implemented in {\scriptsize C}++ that is used for the global minimum of a cost function $cf(\boldsymbol{\theta})$ over the space of phase shifts $\boldsymbol{\theta}$. In the case of universal multiport interferometers, the algorithm was run independently for every value of $\alpha$ and it explored the space of phase shifts $\boldsymbol{\theta}=(\boldsymbol{\theta}',\boldsymbol{\theta}'',\boldsymbol{\theta}''')$ of all beam-splitters. We used the following cost function:
    \begin{equation}\label{eqn:cost_function}
        cf^{(1)}_{\alpha}(\boldsymbol{\theta})=-\tilde{p}^{\mu}(\boldsymbol{\theta})F(\alpha,\boldsymbol{\theta}),
    \end{equation}
where $\tilde{p}(\boldsymbol{\theta})$ is the probability of obtaining a state with $|\mathbf{d}\rangle_a$ in the auxiliary modes calculated as $\tilde{p}(\boldsymbol{\theta})=\langle\chi_{\mathbf{d}}(\boldsymbol{\theta})|\chi_{\mathbf{d}}(\boldsymbol{\theta})\rangle$, where $|\chi_{\mathbf{d}}(\boldsymbol{\theta})\rangle=_a\langle\mathbf{d}|\tilde{\Psi}(\boldsymbol{\theta})\rangle_{sa}$ is the projection of the prepared state $|\tilde{\Psi}(\boldsymbol{\theta})\rangle_{sa}$ on the heralding pattern and $F(\alpha,\boldsymbol{\theta})=|\langle\chi_{\mathbf{d}}(\boldsymbol{\theta})|\Phi(\alpha)\rangle|^2/\tilde{p}(\boldsymbol{\theta})$ is the fidelity of the projected state $|\chi_{\mathbf{d}}(\boldsymbol{\theta})\rangle$ with respect to the target $|\Phi(\alpha)\rangle$. $\mu$ is chosen for better convergence of the algorithm~\cite{fldzhyan2021compact}. In a successive run, the cost function should achieve its global minimum at a set of phase shifts $\boldsymbol{\theta}^{*}=\boldsymbol{\theta}^{*}(\alpha)$ corresponding to both an ideal state with $F(\alpha,\boldsymbol{\theta}^{*})=1$ and maximum probability $p(\alpha)=\tilde{p}(\boldsymbol{\theta}^{*})$. 

The cost function given in Eq.~\eqref{eqn:cost_function} necessitates the calculation of the output-state vector $|\tilde{\Psi}(\boldsymbol{\theta})\rangle_{sa}$ each time the interferometer's phase shifts $\boldsymbol{\theta}$ are updated by the optimization algorithm. For this purpose, the state at the output of a lossless $N$-mode interferometer is expressed as a superposition $|\tilde{\Psi}\rangle_{sa}=\sum_{\mathbf{t}}c_{\mathbf{t},\mathbf{s}}|\mathbf{t}\rangle_{sa}$ over the basis states $|\mathbf{t}\rangle_{sa}$, where  $c_{\mathbf{t},\mathbf{s}}$ are the probability amplitudes specifying a state. We denote the states and the probability amplitudes by the photon occupation vectors $\mathbf{s}=(s_1,\ldots,s_N)$ and $\mathbf{t}=(t_1,\ldots,t_N)$ at the input and the output, respectively. In general, the sum runs through all possible vectors $\mathbf{t}$, the number of which scales as $\begin{pmatrix}
  M+N-1\\ 
  M
\end{pmatrix}$, where $M$ is the number of photons and $N$ is the number of modes. Following the assignment of modes adopted in Fig.~\ref{fig:fig_1}, the basis states are split into a logical ``s" and an auxiliary ``a" part: $\mathbf{t}=\mathbf{t}_s\oplus\mathbf{t}_a$, so $|\mathbf{t}\rangle_{sa}=|\mathbf{t}_s\rangle_{s}|\mathbf{t}_a\rangle_{a}$. The optimization program calculates the probability amplitudes as  $c_{\mathbf{t},\mathbf{s}}=\text{perm}(U_{\mathbf{t},\mathbf{s}})/\sqrt{\mathbf{t}!\cdot\mathbf{s}!}$, where  $\text{perm}(U_{\mathbf{t},\mathbf{s}})$ is a permanent of the matrix $U_{\mathbf{t},\mathbf{s}}$. The matrix $U_{\mathbf{t},\mathbf{s}}$ is derived from the interferometer transfer matrix $U$ by taking the columns and rows according to vectors $\mathbf{s}$ and $\mathbf{t}$, respectively \cite{Tichy}; $\mathbf{s}!=s_1!\times\ldots\times s_N!$ and $\mathbf{t}!=t_1!\times\ldots\times t_N!$ are shorthand for factorials.

\subsection{Restricted linear optics}

The drawback of using universal $N$-port interferometers is the necessity to program a large set of parameters $\boldsymbol{\theta}=(\boldsymbol{\theta}',\boldsymbol{\theta}'',\boldsymbol{\theta}''')$ consisting of $N^2-1$ phase shifts every time a state with a new $\alpha$ is required. For quantum photonic circuits generating a narrow class of states depending on only one parameter, this can be impractical. More-compact circuits reprogrammed by a smaller number of phase shifts, if they exist, would simplify experimental realizations by making the photonic circuit smaller and power efficient, and by simplifying the reprogramming procedure. The question of whether compact schemes exists is one that we are answering in our work.

    \begin{figure}[htp]
        \centering
        \includegraphics[width=0.4\textwidth]{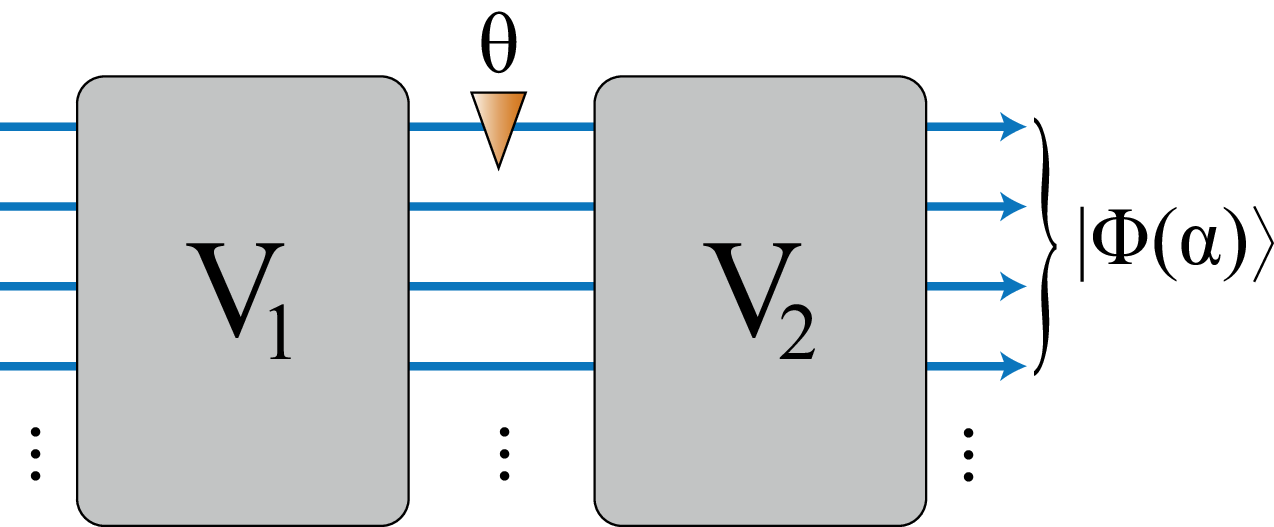}
        \caption{Compact programmable interferometer with one variable phase shift $\theta$ considered in the restricted design for the generation of two-qubit states $|\Phi(\alpha)\rangle$. The  phase shift $\theta$ is sandwiched between two static interferometers with transfer matrices $V_1$ and $V_2$. }
        \label{fig:fig_4}
    \end{figure}

\subsubsection{Circuits with one programmable phase}

Since the states of interest depend on one parameter, we  were interested in schemes reprogrammed by only one phase shift for all values of $\alpha$, which is shown in Fig.~\ref{fig:fig_4}. The phase shift $\theta$, by means of which a specific generated state is programmed, is sandwiched between two static interferometers, which we  designate by corresponding transfer matrices $V_1$ and $V_2$. 

To find the unknown static transfer matrices $V_1$ and $V_2$, an optimization algorithm can be used as before, however, with a new cost function $cf^{(2)}$ accounting for the constraints of the restricted programmable circuit. In the construction of the cost function we parametrize each matrix $V_1$ and $V_2$ by the meshes of beam-splitters, similarly to the universal design, but with fewer layers, $L$,  than required for universality, starting with $L=2$ (see Appendix \ref{appendix:parametrization} for details). This can ease the numerical optimization and make the resultant circuit more compact. The optimization explores the parameter space of the following interferometer transfer matrix:
    \begin{equation}
        U(\theta,\boldsymbol{\psi}^{(1)},\boldsymbol{\psi}^{(2)})=V_2(\boldsymbol{\psi}^{(2)})\text{diag}(e^{i\theta},1,\ldots,1){}V_1(\boldsymbol{\psi}^{(1)}),\label{eqn:compact_circuit}
    \end{equation}
where $\boldsymbol{\psi}^{(1)}$ and $\boldsymbol{\psi}^{(2)}$ are parameters of the static beam-splitter meshes  $V_1$ and $V_2$. Notice that  parameters $\boldsymbol{\psi}=(\boldsymbol{\psi}^{(1)},\boldsymbol{\psi}^{(2)})$ are used only for optimization and they are not changed when one is reprogramming the generated state $|\Phi(\alpha)\rangle$ by the variable phase shift $\theta$. 

The cost function that catches the constraints of the interferometer shown in Fig.~\ref{fig:fig_4} should quantify the generation errors simultaneously for all states $|\Phi(\alpha)\rangle$. However, in numerical analysis this boils down to calculating the error for a discrete set of states  $|\Phi(\alpha_m)\rangle$, where we assume the ordering: $0\le\alpha_1<\cdots<\alpha_m<\cdots<\alpha_S\le\pi/4$. Thus, the cost function to find the quantum circuit with one programmable phase shift is given by
    \begin{equation}\label{eqn:cost_function_2}
        cf^{(2)}(\{\theta_m\},\boldsymbol{\psi})=\frac{1}{S}\sum_{m=1}^Scf^{(1)}_{\alpha_m}(\theta_m,\boldsymbol{\psi}),
    \end{equation}
which uses the definition of the cost function given in Eq.~\eqref{eqn:cost_function}. In Eq.~\eqref{eqn:cost_function_2} there are two different subsets of parameters---$\boldsymbol{\psi}$ and $\{\theta_m\}$.  The set $\{\theta_m\}$ corresponds to the values of $\theta$ to generate respective states $\{|\Phi(\alpha_m)\rangle\}$ in the optimal quantum circuit. The set $\boldsymbol{\psi}$ defines the static part of the quantum circuit, which is independent of the target state $|\Phi(\alpha)\rangle$.


\subsubsection{Compactifying static linear optics}

Besides using static low-depth interferometers, one can make them even more compact and simple for practical implementations by building them from the minimal number of beam-splitters and (static) phase shifts. To this end, we used yet another cost function, which is essentially a modification of Eq.~\eqref{eqn:cost_function_2}: 
    \begin{eqnarray}\label{eqn:cost_function_3}
        &cf^{(3)}(\{\theta_m\},\boldsymbol{\psi})=\\\nonumber
        &\frac{1}{S}\sum_{m=1}^S(cf^{(1)}_{\alpha_m}(\theta_m,\boldsymbol{\psi})+\frac{\varepsilon}{Q}\sum_{q=1}^Q(\sin^22\psi_q'+\sin^22\psi_q'')).
    \end{eqnarray}
Here, an extra term $\frac{\varepsilon}{Q}\sum_{q=1}^Q(\sin^22\psi_q'+\sin^22\psi_q'')$ is added to the summands in Eq.~\eqref{eqn:cost_function_2}, where phase shifts $\psi_q'$ and $\psi_q''$ quantify the beam-splitter parameters in the static meshes, and $Q$ is the total number of parameters $\boldsymbol{\psi}$ in $V_1$ and $V_2$ combined. This extra term can force the cost function to find the configurations with a lesser number of nontrivial beam-splitters~\cite{fldzhyan2021compact,gubarev2020}, since its decreases towards transmissivities $\tau_q=\cos^2\psi_q'$ are either $0$ or $1$ and phase shifts $\psi_q''$ either $0$ or $\pi/2$.

In Eq.~\eqref{eqn:cost_function_3}, $\varepsilon$ controls the interplay between the terms $cf^{(1)}_{\alpha}(\theta,\boldsymbol{\psi})$ that tend to minimize infidelity and increase the success probability and the terms that tend to make static interferometers more compact. Obviously, large values of $\varepsilon$ can ruin the fidelity of success probability of the linear optical quantum circuit---its two main characteristics. Therefore, its value, which is set before the optimization algorithm is run, should be chosen with special care. We present the corresponding analysis in Appendix \ref{appendix:analysis_varepsilon}.

    \begin{figure}[htp]
        \centering
        \includegraphics[width=0.45\textwidth]{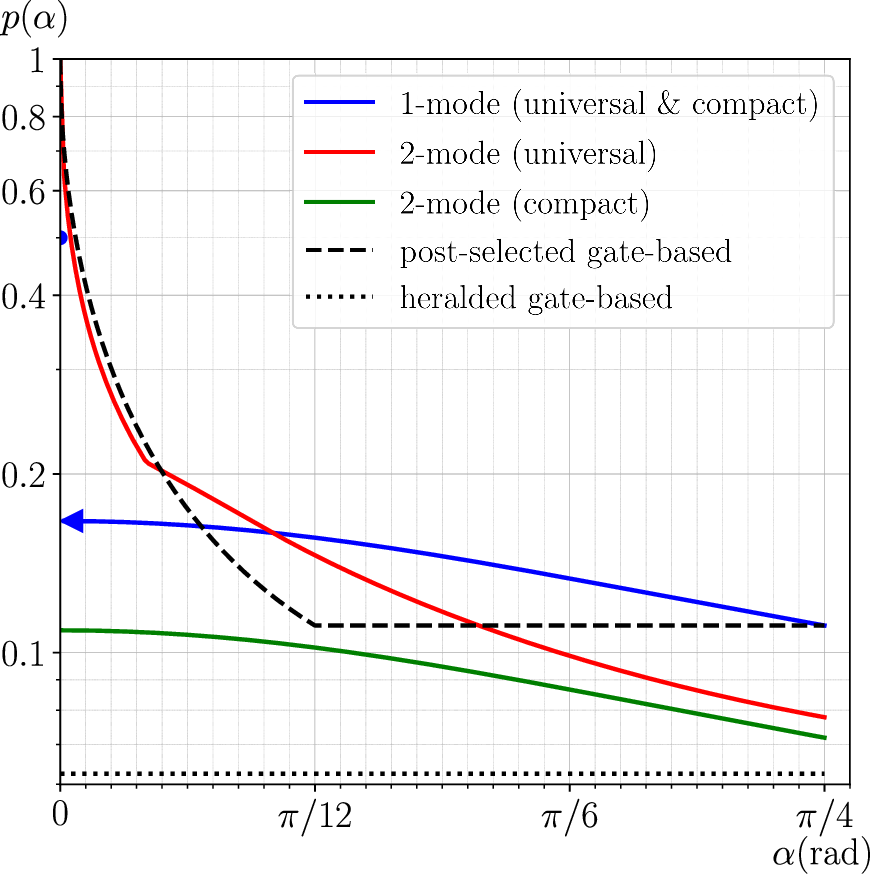}
        \caption{Success probability $p$ as a function of state parameter $\alpha$ in the cases of universal (all phases are variable) and compact (one variable phase) interferometers for schemes with two-mode  and single-mode heralding. The curve for the compact two-mode-heralded scheme is obtained by minimization of the cost function given by Eq.~\eqref{eqn:cost_function_2} at $S=3$ points: $\alpha_1=\pi/60$, $\alpha_2=5\pi/72$, and $\alpha_3=\pi/4$ ($\varepsilon=2\times10^{-3}$ and $\mu=10^{-5}$). The dashed and dotted curves correspond to the optimal post-selected and heralded gate--based linear optical circuits with gate success probabilities given by Eq.~\eqref{eqn:cphase_dependence} and $1/16$, respectively (see Fig.~\ref{fig:fig_3}). }
        \label{fig:fig_5}
    \end{figure}
    \begin{figure}[htp]
        \centering
        \includegraphics[width=0.45\textwidth]{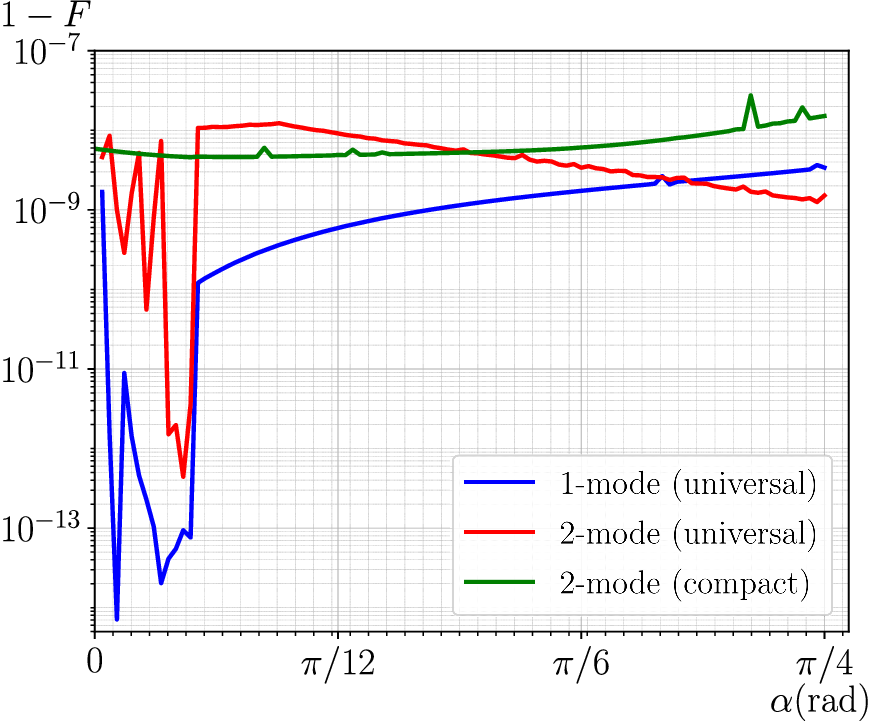}

        \caption{Infidelity $1-F$ as a function of the parameter $\alpha$ that accompanies the probability dependencies depicted in Fig.~\ref{fig:fig_5} in the course of numerical optimization. The parameters are the same as in Fig.~\ref{fig:fig_5}.}
        \label{fig:fig_6}
    \end{figure}

The search for a quantum optical scheme by  $cf^{(3)}(\{\theta_m\},\boldsymbol{\psi})$ starts with choosing the interferometer depths $L$ and $S$ values of $\alpha_m$. Then, the optimization algorithm searches for the global minima of $cf^{(3)}(\{\theta_m\},\boldsymbol{\psi})$ simultaneously with respect to both $\{\theta_m\}$ and $\boldsymbol{\psi}$. If the fidelity obtained is near perfect and success probability is high enough, the quantum circuit is considered found. Otherwise, the depth $L$ is increased and the algorithm runs again. Since with $L=N$ one can obtain arbitrary unitary transfer matrices $V_1$ and $V_2$ permitted by linear optics, this is the maximum depth beyond which static linear optics was not considered.

\section{Results}\label{sec:results}

\subsection{Universal linear optics}

Fig.~\ref{fig:fig_5} and Fig.~\ref{fig:fig_6} show the dependencies of the probability $\tilde{p}$ and the infidelity $1-F$, respectively, on $\alpha$ that were obtained by numerical optimization for the two-mode-heralded and the single-mode-heralded scheme using universal and compact interferometers. The probability curves describing the linear optical gate--based circuits are also plotted in Fig.~\ref{fig:fig_5}. To prevent the algorithm from sticking to a local minimum and to guarantee convergence of the loss function to the global minimum, it was run several times (typically three to five times), each time with our initializing the phase shifts with random values. The numerical algorithm does not get to exactly zero infidelity as typically is the case with numerical algorithms; however, we attribute the small infidelities on the order of approximately $10^{-10}-10^{-8}$ in Fig.~\ref{fig:fig_6} to ideal target states with $1-F=0$.

First, we searched for schemes based on universal interferometers capable of exploring the full space of transformations enabled by linear optics using the cost function $cf^{(1)}_{\alpha}(\boldsymbol{\theta})$ \eqref{eqn:cost_function}. As can be seen from Fig.~\ref{fig:fig_5}, for both the two-mode-heralded and the single-mode-heralded scheme, decrease of the entanglement degree of the target state (quantified by the value of $\alpha$) gives higher efficiency. However, the dependencies behave in a markedly different way for two-mode heralding and single-mode heralding. The scheme with single-mode heralding is more efficient in generating highly entangled states than the scheme with two-mode heralding: its probability curve starts from $p=1/9$ for the Bell state $|\Phi(\alpha=\pi/4)\rangle$ and has a maximal value of $p=1/6$ when approaching the separable state $|\Phi(\alpha=0)\rangle=|0101\rangle$. Notice that this probability is different from the $1/2$ obtained for the universal interferometer at $\alpha=0$. Also notice that this state cannot be generated with unit efficiency, as one might expect for the separable state $|\Phi(0)\rangle$, which is explained by the presence of the two-photon Fock state $|2\rangle_a$ in the auxiliary mode; obviously, it is impossible to concentrate two single photons in a single mode with probability exceeding $1/2$ by linear optics \cite{Kieling_PhD}.

Use of two-mode heralding is better if one needs states with a low entanglement degree. Moreover, at $\alpha\ll{}1$, use of two-mode heralding is much better  as its probability approaches $1$ in striking contrast to the one-mode-heralded scheme. The crossover point for the two probability curves is located at $\alpha\approx{0.22}$ rad, which corresponds to the qubit state $|\Phi_x\rangle=0.976|0101\rangle+0.218|1010\rangle$. Unfortunately, numerical analysis does not give a physical explanation for this remarkable phenomenon; therefore, it is left for further research. Fig.~\ref{fig:fig_5} suggests that our specifically designed heralded linear optical scheme is more efficient than the most-effective (but nonscalable) post-selected gate--based approach known today in almost all the parameter range. When compared with the known heralded (scalable) gate--based approach, our scheme is much more effective in generating all the parameter range of the entangled states.

\subsection{Restricted linear optics}

\subsubsection{Compact scheme with single-mode heralding}

We studied the restricted compact interferometers that have one variable phase shift as shown in Fig.~\ref{fig:fig_4}. For this, the cost function  $cf^{(2)}(\{\theta_m\},\boldsymbol{\psi})$ was used. It was found that the dependence of $p(\alpha)$ for the five-port programmable scheme obtained by our assuming universal linear optics (cost function $cf^{(1)}_{\alpha}(\boldsymbol{\theta})$) completely coincides with the dependence for restricted linear optics (cost function $cf^{(2)}_{\alpha}(\{\theta_m\},\boldsymbol{\psi})$). In other words, the compact five-port scheme with single-mode heralding enables the use of only one programmable phase shift without sacrificing success probability and fidelity. To search for an even-more-compact and even-simpler scheme consisting of the minimal number of nontrivial static elements, we used the cost function $cf^{(3)}(\{\theta_m\},\boldsymbol{\psi})$. The corresponding probability curve in Fig.~\ref{fig:fig_5} coincides exactly with the curves obtained by our optimizing $cf^{(1)}(\boldsymbol{\theta})$ and $cf^{(2)}(\{\theta_m\},\boldsymbol{\psi})$.
 
The set of optimal parameters $\boldsymbol{\psi}$ obtained by the numerical algorithm typically contained some phase shift parameters that did not equal the trivial values, but that were close to them. After computer optimization these values were ``rounded" to the corresponding closest trivial values, and then the state fidelity and success probability were calculated again, so as to validate the derived optimal circuit.

\begin{figure}[htp]
        \centering
        \includegraphics[width=0.45\textwidth]{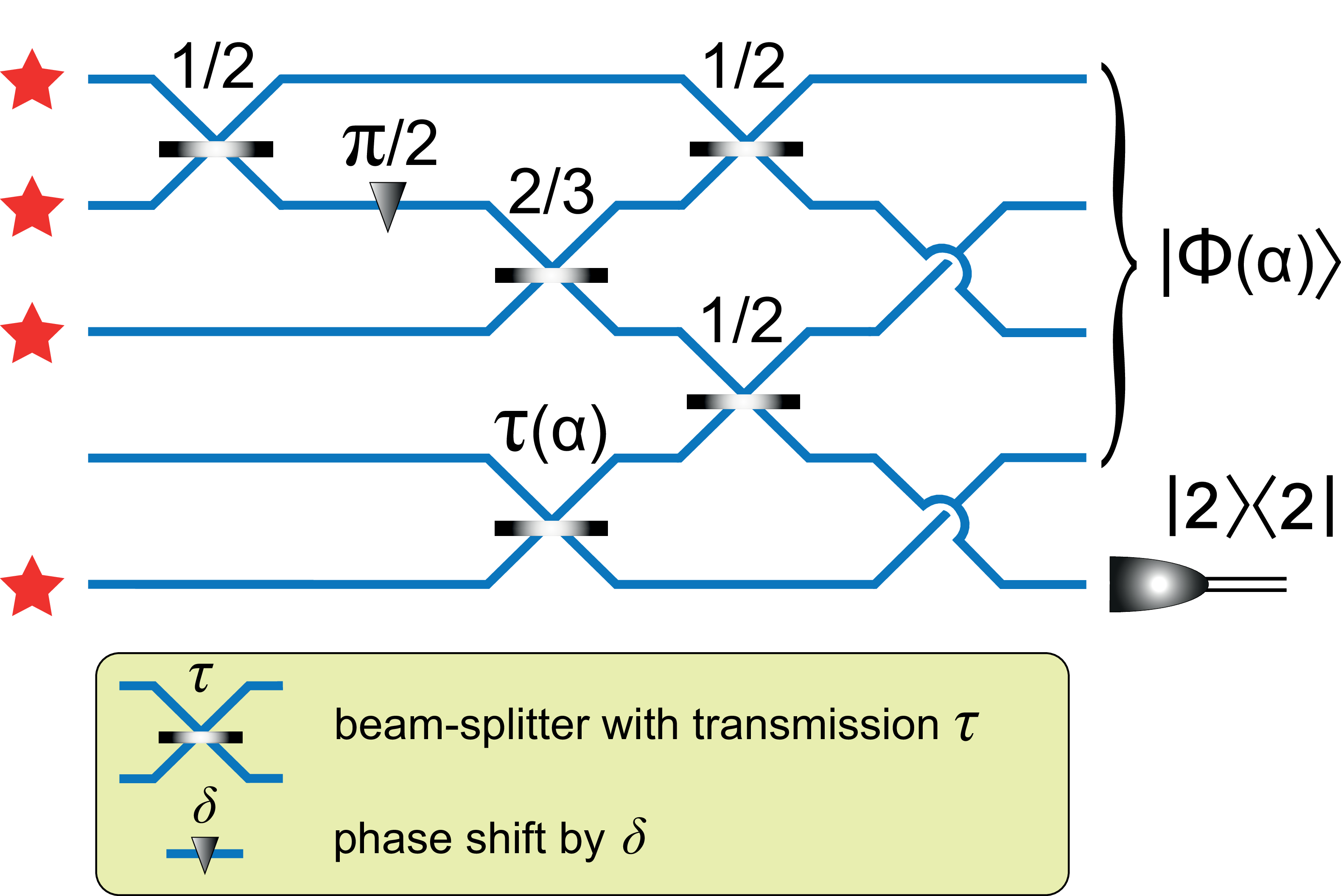}
        \caption{Explicit form of the compact five-mode programmable interferometer generating two-qubit states $|\Phi(\alpha)\rangle$. The beam-splitters are marked by their power transmissivities $\tau_j$ ($\varphi_j=0$ for all beam-splitters). A state with a specified value of $\alpha$ is generated by the setting of the transmission of the bottom beam-splitter $\tau(\alpha)$ to $1/(1+2\tan^2\alpha)$.}
        \label{fig:fig_7}
    \end{figure}

The resultant layout of the minimal scheme found is shown in Fig.~\ref{fig:fig_7}.
It is built up of five static beam-splitters, one static phase shift, and one variable beam-splitter with transmissivity $\tau(\alpha)$. The latter is essentially a Mach-Zehnder interferometer consisting of two static beam-splitters. Therefore, in total, our scheme is built up of $Q_{\text{nontriv}}=7$ static elements (see Fig.~\ref{fig:fig_9} in Appendix \ref{appendix:analysis_varepsilon}). 

For this minimal scheme, one can derive the output state \eqref{eqn:optimal_state_form} and the generated heralded state analytically. This gives the success probability as
    \begin{equation}
        p(\alpha)=\frac{1}{6(1+\sin^2\alpha)},\quad(0\le\alpha\le\pi/4),
    \end{equation}
and the beam-splitter transmissivity to be set in the quantum circuit to generate a state $|\Phi(\alpha)\rangle$ as
    \begin{equation}
        \tau(\alpha)=\frac{1}{1+2\tan^2\alpha}.
    \end{equation}
It turns out that the variable beam-splitter is the only  modification of the  scheme for Bell-state generation found previously ~\cite{fldzhyan2021compact}. 
  
This finding makes unnecessary the use of complex interferometers with multiple programmable phase shifts to generate two-qubit states \eqref{eqn:parametrized_state} by single-mode heralding.

\subsubsection{Compact scheme with two-mode heralding}

Similarly, we tried to find the minimal decomposition of the scheme that uses two-mode heralding. To this end, we used the cost function $cf^{(2)}(\boldsymbol{\psi})$ constructed with a different number $S$ of points $\alpha_m$ distributed in the range from $0$ to $\pi/4$. However, we did not find the minimal circuit that would generate states $|\Phi(\alpha)\rangle$ with maximal probability $p(\alpha)$ that was obtained with universal interferometers. Fig.~\ref{fig:fig_5} and Fig.~\ref{fig:fig_6} show probability and infidelity curves, respectively, for one example in which $S=3$ points were used. As can be seen from  Fig.~\ref{fig:fig_5}, the probability is much lower than optimal, as achieved by universal linear optics using the same heralding measurement pattern.

To get a deeper understanding of the negative results for the existence of a two-mode-heralded scheme with one programmable phase shift, we analyze the transfer matrices of the universal interferometers in Appendix \ref{sec:svd_analysis}. The analysis suggests the success-probability dependence in the case of the six-port universal scheme, shown in Fig.~\ref{fig:fig_5}, is associated with at least three different programmable quantum circuits, each of which provides optimal generation in a certain range of the target-state parameter $\alpha$.

\section{Conclusion}\label{sec:discussion}

We studied the limits of linear optics in the generation of arbitrary dual-rail-encoded two-qubit states using four photons, which is the minimal possible number for heralded generation of these states. Besides applications in fundamental research\cite{Gomez_2019,Ishizaka_2020}, such states can be of interest for generic multiqubit-resource states, in particular, weighted-graph states~\cite{our_future_work}.

Two schemes, one using single-mode heralding and the other using two-mode heralding, which are the most-compact possible, were considered. It was shown that the success probability for both schemes increases when the target-state entanglement degree increases, however following essentially different dependencies. The scheme with single-mode heralding starts with a success probability of $1/9$ for the maximally entangled Bell state and increases to $1/6$ in the limit of separable states. The scheme with two-mode heralding, while having lower success probability for highly entangled states---in particular, for Bell states the success probability is $2/27$~\cite{fldzhyan2021compact} (Gubarev showed that it can be slightly higher~\cite{Gubarev2021}) is much-more effective for the generation of weakly entangled states, where the success probability tends to unity in the limit of separable states. The crossover point of the two probability dependencies is found to be at  $|\Phi(\alpha\approx{}0.22)\rangle=0.976|0101\rangle+0.218|1010\rangle$. Unfortunately, because of the lack of analytical theory, we could not gain deep insight into the physical explanation for the occurrence of this discovery, so it is left for future research.

In addition, we showed that the optimal scheme with single-mode heralding can be compactly constructed with use of only one variable phase shift element. However, a similar compact decomposition for the scheme with two-mode heralding was not found. We understand that this is by no means solid evidence for the absence of such a decomposition.

The schemes studied are not immune to errors associated with imperfections of components, such as a source's inefficiencies and a photon's partial distinguishability and photon loss, as using the heralding measurement outcomes, one cannot rule out these cases. As a result of these imperfections, errors degrading the fidelity of the target state are introduced or false-positive generation events occur when the state is far from the target due to a lost photon or lost photons. 
In recent years, significant progress has been made in relevant photonic technologies---high-quality photon sources~\cite{Paesani2020,Tomm2021}, efficient photon detectors~\cite{UnitEfficiencyDetector,TES} and low-loss programmable multiport interferometers~\cite{QuiX1550,QuiX920}, as well as partial integration of the disparate components on a single integrated photonic chip~\cite{Paesani2020,SourceIntegrated1,SourceIntegrated2,SNSPDIntegrated,TESintegrated}. This makes us believe that linear optical methods are a viable approach to implementing quantum information processing.

\section{Acknowledgments}

The authors acknowledge partial support by the Interdisciplinary Scientific and Educational School of Moscow University ''Photonic and Quantum Technologies. Digital Medicine'' and the  Russian Foundation for Basic Research (RFBR Project No. 19-52-80034). S.A.F. and M.Yu.S. are grateful to the Foundation for the Advancement of Theoretical Physics and Mathematics (BASIS)  (Projects No. 20-2-1-97-1 and No. 20-1-3-31-1) and the Russian Roadmap on Quantum Computing for support. 

The authors declare no conflicts of interest.

\appendix

\section{Parametrization of the multiport interferometers }\label{appendix:parametrization}

\begin{figure}[htp]
        \centering
        \includegraphics[width=0.35\textwidth]{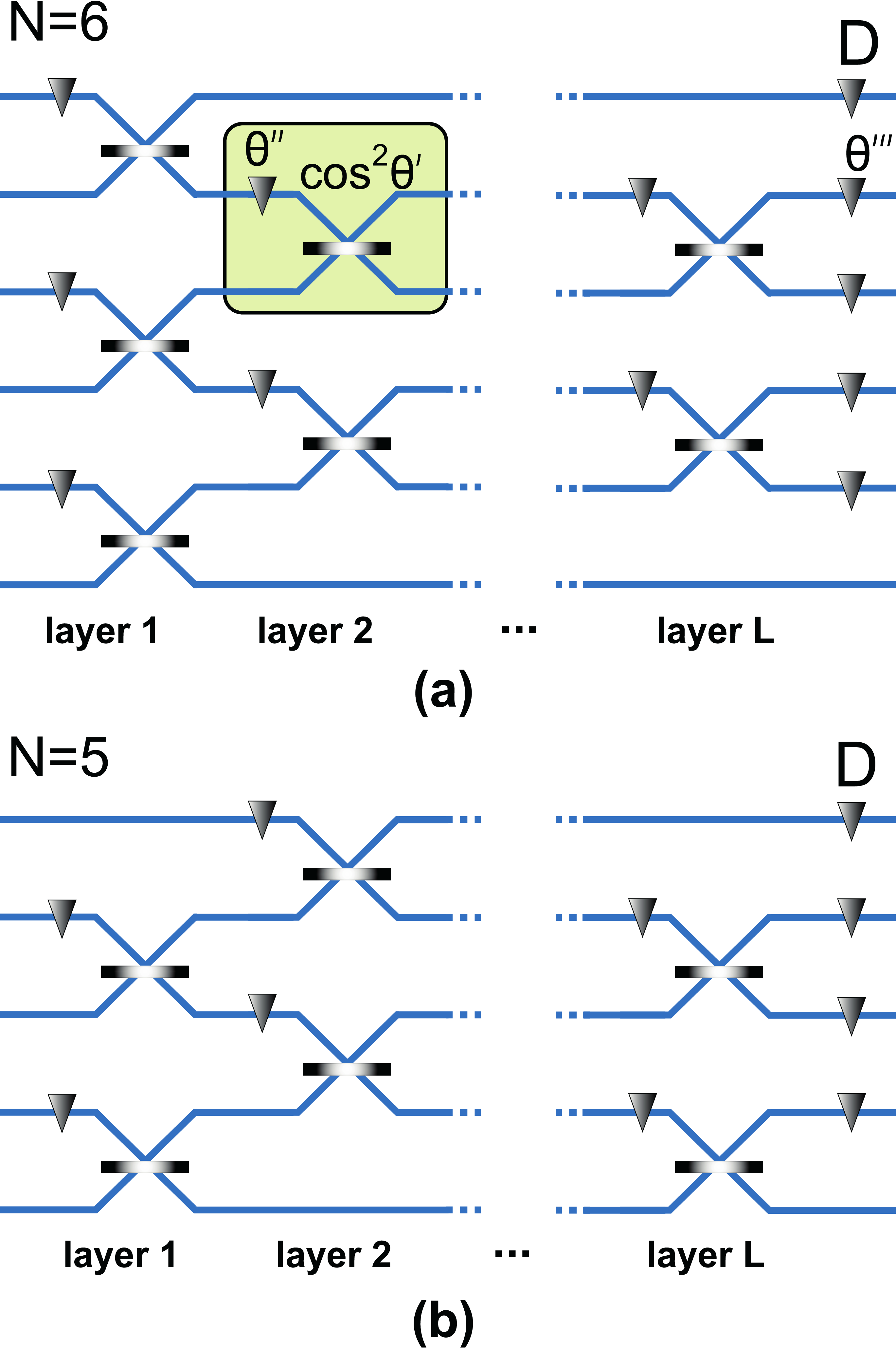}
        \caption{Implementations of programmable $N$-mode interferometers as meshes of variable two-mode beam-splitters and phase shifts for (a) $N=6$ and (b) $N=5$. The highlighted programmable block corresponds to the transfer matrix \eqref{eqn:BSmatrix}. }
        \label{fig:fig_8}
    \end{figure}

To parametrize the transfer matrices of the programmable interferometers ($U^{(5)}(\boldsymbol{\theta})$ and $U^{(6)}(\boldsymbol{\theta})$ in Fig.~\ref{fig:fig_1}) and static multiport interferometers (V$_1(\boldsymbol{\psi}^{(1)})$ and V$_2(\boldsymbol{\psi}^{(2)})$ in Fig.~\ref{fig:fig_4}) we used the beam-splitter meshes shown in Fig.~\ref{fig:fig_8} (the designation of the static phase shifts $\boldsymbol{\psi}^{(1)}$ and $\boldsymbol{\psi}^{(2)}$ are intentionally different from those of programmable ones $\boldsymbol{\theta}$ to avoid confusion). These meshes are built up of programmable blocks, one of which is highlighted in Fig.~\ref{fig:fig_8}(a). The transfer matrix of the highlighted block is given by Eq.~\eqref{eqn:BSmatrix}. In addition to the two-port blocks, extra phase shifts are placed at the output of the meshes described by the transfer matrix $D$. 

The meshes with $L=N$ correspond to the universal design by Clements \textit{et. al.}~\cite{Clements2016} with $N^2-1$ programmable phase shifts. This universal design was used to study the capabilities of universal linear optics in the generation of the states. 

The static interferometers V$_1(\boldsymbol{\psi}^{(1)})$ and V$_2(\boldsymbol{\psi}^{(2)})$ have $L\le{}N$. The search for the most-compact static interferometers started from optimizations with $L=2$. If success was not achieved in the course of the optimization, one more layer was added and the optimization was started again

\section{Optimal values of $\varepsilon$ in the cost function $cf^{(3)}$ }\label{appendix:analysis_varepsilon}

To find the range of $\varepsilon$ in which one can find a minimal quantum circuit with one variable phase shift (see Fig.~\ref{fig:fig_4}), we analyzed the dependence of the cost function $cf^{(3)}$ on $\varepsilon$. We were specifically interested in the number of nontrivial beam-splitters and phase shifts, $Q_{\mathrm{nontriv}}$, which was obtained with the use of the cost function given by Eq.~\eqref{eqn:cost_function_3} with a given value of $\varepsilon$. In the numerical simulation, a beam-splitter or a phase shift was considered nontrivial if the following conditions were fulfilled:
    \begin{eqnarray}
        &0.001<\cos^2\psi_q'< 0.999,\text{ for beam-splitters}\\ \nonumber
        &\cos\psi_q''< \sqrt{0.999},\text{ for phase-shifters}.
    \end{eqnarray}
Such elements are counted in $Q_{\mathrm{nontriv}}$. Otherwise, an element is considered trivial and in the resultant circuit $\cos\psi_q'$ or  $\cos\psi_q''$ is rounded to either $0$ or $1$ depending on the closest value.

\begin{figure}[htp]
        \centering
        \includegraphics[width=0.45\textwidth]{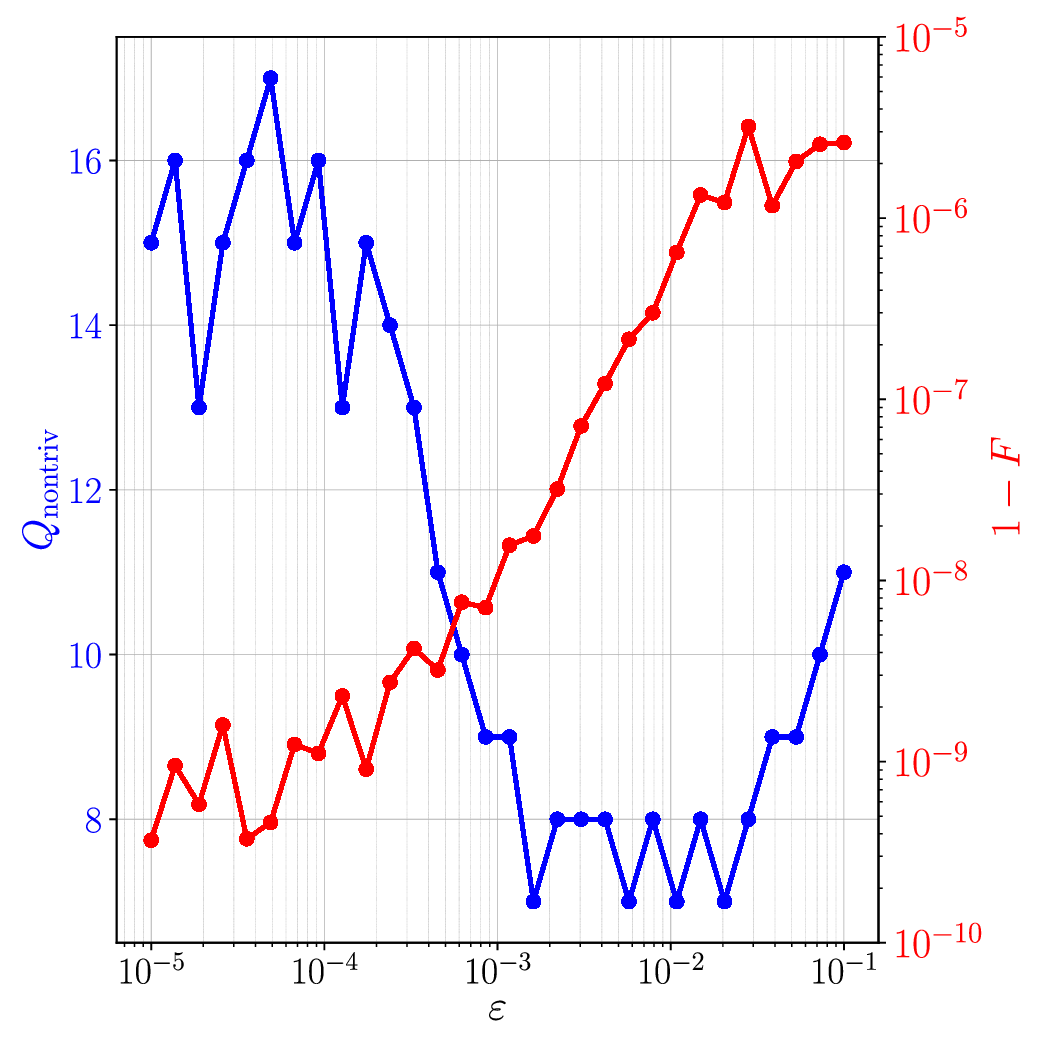}
        \caption{Dependence of the number of nontrivial static elements in the compact five-port scheme and generated state infidelity on the parameter $\varepsilon$ in the cost function given by Eq.~\eqref{eqn:cost_function_3}. The numerical simulation was performed at interferometer depths $L_1=L_2=3$ and for $\mu=10^{-5}$.}
        \label{fig:fig_9}
    \end{figure}

Fig.~\ref{fig:fig_9} shows the dependence of $Q_{\mathrm{nontriv}}$ and the generated state infidelity $1-F$ on $\varepsilon$ obtained by our optimizing the cost function $cf^{(3)}$ for the target five-mode scheme with $S=4$ trial states $\alpha_1=\pi/16$,  $\alpha_2=\pi/8$, $\alpha_3=3\pi/16$, and $\alpha_4=\pi/4$. For each value of $\varepsilon$, the optimization was repeated $100$ times and the minimum found with the lowest $Q_{\mathrm{nontriv}}$ was chosen for visual presentation. As can be seen from the Fig.~\ref{fig:fig_9}, there is a range of $\varepsilon$ values approximately $10^{-3}$ to approximately $2\cdot10^{-2}$ that yield minimal $Q_{\mathrm{nontriv}}=7$ or $Q_{\mathrm{nontriv}}=8$. In the numerical simulations, the value $\varepsilon=2\times 10^{-3}$, which worked well in obtaining the compact scheme depicted in Fig.~\ref{fig:fig_7}.

\section{Singular-eigenvalue analysis of the universal linear optics}\label{sec:svd_analysis}

The restricted single-parameter $\theta(\alpha)$ decomposition in Fig~\ref{fig:fig_4} reasonably implies that all properties of the optimal generation scheme are smooth functions of $\alpha$. However, as we show here, this is not the case for a six-port scheme.

The state \eqref{eqn:optimal_state_form} at the output of the optimized six-port interferometer can be written in the form
    \begin{equation}
        |\Psi(\alpha)\rangle=\text{perm}(U_{\boldsymbol{t}_1,\boldsymbol{s}}^{(6)}(\alpha))|\boldsymbol{t}_1\rangle+\text{perm}(U_{\boldsymbol{t}_2,\boldsymbol{s}}^{(6)}(\alpha))|\boldsymbol{t}_2\rangle+\cdots
    \end{equation}
where $\boldsymbol{t}_1=(1,0,1,0,1,1)$ and $\boldsymbol{t}_2=(0,1,0,1,1,1)$ are the occupation photon-number vectors designating the basis states at the output of the interferometer relevant for the generation of the Bell state, and $\boldsymbol{s}=(1,1,1,1,0,0)$ is the input occupation photon-number vector. 

\begin{figure}[htp]
        \centering
        \includegraphics[width=0.45\textwidth]{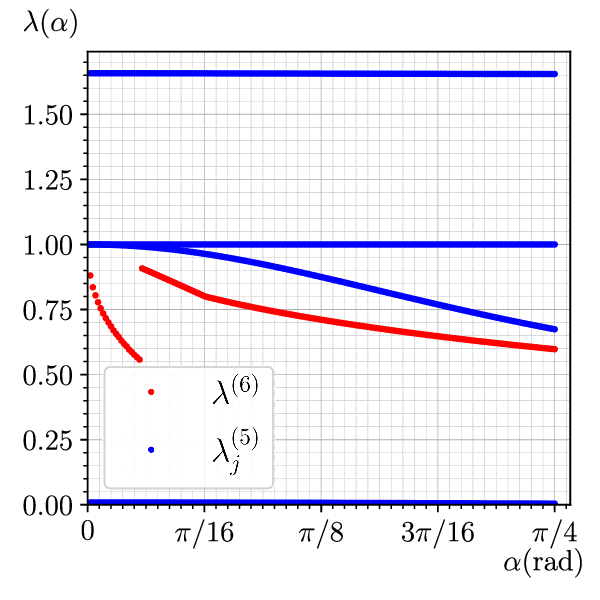}
        \caption{The dependence of the singular eigenvalue $\lambda^{(6)}$ of submatrix $U_{\boldsymbol{t}_1,\boldsymbol{s}}^{(6)}(\alpha)$ and singular eigenvalues $\lambda^{(5)}_j$ ($j=\overline{1,4}$) of submatrix $U_{\boldsymbol{t}_1,\boldsymbol{s}}^{(5)}(\alpha)$ corresponding to the dependencies shown in Fig.~\ref{fig:fig_5}.}
        \label{fig:fig_10}
    \end{figure}

To show that the dependence of $p(\alpha)$ for the universal two-mode-heralded scheme, shown in Fig.~\ref{fig:fig_5}, consists of multiple subcurves corresponding to different linear optical circuits, we analyzed the singular values of submatrices $U_{\boldsymbol{t}_1,\boldsymbol{s}}^{(6)}(\alpha)$ and $U_{\boldsymbol{t}_2,\boldsymbol{s}}^{(6)}(\alpha)$. It turns out that the singular-value spectrum of submatrix $U_{\boldsymbol{t}_1,\boldsymbol{s}}^{(6)}(\alpha)$ contains only one eigenvalue $\lambda^{(6)}(\alpha)$ that differs from $1$: $(\lambda^{(6)}(\alpha),\lambda^{(6)}(\alpha),1,1)$. Fig.~\ref{fig:fig_10} shows the dependence of $\lambda^{(6)}(\alpha)$, which corresponds to the dependency of $p(\alpha)$ shown in Fig.~\ref{fig:fig_5} for universal linear optics. As can be seen, the curve has a break at $\alpha\approx\pi/16$ and a derivative discontinuity at  $\alpha\approx\pi/36$. We attribute this behavior to the existence of at least three families of quantum circuits, each of which is optimal in a certain range of $\alpha$.

Similarly, we plot the singular eigenvalues for the submatrix $U_{\tilde{\boldsymbol{t}}_1,\boldsymbol{s}}^{(5)}(\alpha)$ of the five-port quantum scheme, where $\tilde{\boldsymbol{t}}_1=(1,0,1,0,2)$. In this case, there are four distinct eigenvalues $\lambda^{(5)}_j$ ($j=\overline{1,4}$), which have smooth dependencies (excluding $\alpha=0$), in contrast to $\lambda^{(6)}(\alpha)$. The eigenvalues of the other submatrix $U_{\tilde{\boldsymbol{t}}_2,\boldsymbol{s}}^{(5)}(\alpha)$, where $\tilde{\boldsymbol{t}}_2=(0,1,0,1,2)$, are also smooth without any features.

\bibliography{sorsamp}
\end{document}